# Structure factor realizability reveals the glass-dynamics onset temperature

Bowen Duan[1] and Ge Zhang[1,*]

[1]Department of Physics, City University of Hong Kong, Hong Kong, China

**ABSTRACT**. When liquids are cooled rapidly, they bypass crystallization and instead enter a supercooled state and then a glass state. Previous studies have shown that the static structure factors of high-temperature liquids, supercooled liquids, and glasses exhibit only subtle differences, leading to the conclusion that the glass transition cannot be predicted solely from structure factor changes. Our research challenges this limitation. Specifically, we studied the difficulty of generating configurations corresponding to target structure factors using stochastic gradient descent optimizations. While such optimizations easily converge when targeting the structure factors of higher-temperature liquids, the difficulty significantly increases for lower temperature liquids and glasses. By quantifying this difficulty through the mean squared error achieved, we found a kink at the onset temperature of glass dynamics. Our results suggest that the onset of glass dynamics can be explained by a topological change in the realizability landscape of the structure factor, even though the structure factor itself only changes smoothly. Our results are currently based on computer simulations using original and modified Dzugutov interactions, and future work will determine whether our theory is applicable to other glass-forming systems.

## I. INTRODUCTION

When liquids are cooled rapidly, they bypass crystallization and instead enter a supercooled state and then a glass state. The mechanism of glass transition is one of the most profound and difficult problems in condensed matter physics [1-3]. During the glass transition, the dynamic process of the system is strongly suppressed, resulting in the inability to relax on a laboratory time scale and thus being confined to a non-equilibrium solid state. To understand this drastic increase in the relaxation time, one searches for other quantities that also change drastically or qualitatively as liquids cool.

One fruitful direction of search is in the quantities related to the energy landscape. A study found that the energy of inherent structures (nearest energy minima) is temperature-independent above the glass-dynamics onset temperature, but is temperature-dependent below such temperature [4]. Another study found that the system is closer to a saddle point than a minimum in the energy landscape above the mode-coupling temperature, and vice versa below the mode-coupling temperature [5]. These studies contribute to building a thermodynamic understanding of the glass transition.

There have also been discoveries in energy-independent, purely geometric quantities. By studying the “point-to-set” correlation function, which is a type of many-body correlation function, one discovers a hidden length scale [6], which supports the Random First Order Transition theory of glass transition. Another study utilized machine learning to identify unstable, soft local clusters in glasses, and this leads to an explanation of the super-Arrhenius relaxation time in glasses [7].

While the studies of many-particle statistics have been fruitful [6-7], the pair statistics of glasses have previously been considered uninformative [3]. As the temperature varies, the pair correlation function and the static structure factor change smoothly [3], and no physical insights have been extracted from studying these pair statistics.

However, our study discovers that such smooth changes in the pair statistics actually leads to a qualitative change in their realizability at the onset temperature. Specifically, we calculate the static structure factors of computational glass formers at a series of temperatures, and then use stochastic gradient descent optimization to generate configurations that match these target structure factors. After that, we quantify the differences between the average structure factors of the optimized configurations and the targets by mean squared error. We find that the error is small at high temperatures. However, starting from the onset of glass dynamics, as the temperature decreases, the error begins to significantly increase, indicating that the structure factors become difficult to realize. Thus, the quantitatively small changes in structure factors during the liquid-to-glass transition results in a qualitatively important change in the realizability, which could explain the onset of glass dynamics. As a result, we have also developed an algorithm that only requires static structure factors to identify the onset temperature, which were previously only identifiable using dynamic properties.

## II. METHODS

This study consists of two steps. First, we perform molecular dynamics (MD) simulations to calculate static structure factors at various temperatures. Second, we perform optimizations to generate configurations targeting such structure factors. We study two different single-component systems with original and modified Dzugutov interactions [8,9], respectively.

### A. Calculating structure factors

To calculate the structure factor at various temperatures while excluding crystallization, we designed the following steps:

1. Perform MD simulations at a high temperature and then exponentially cool to a target temperature and sample configurations.
2. Perform energy minimization and calculate local bond order parameter $q_8$ [10] to detect crystallization.
3. If crystallization is not detected, the sampled configurations are used to calculate the structure factors. Otherwise, the entire trajectory is discarded.

We elaborate on each step below.

We use HOOMD [11] to perform MD simulations. Our system consists of $N$=1000 particles in a cubic box with density $\rho$=0.87. The density is chosen to be consistent with a previous research [12]. The particles interact with original Dzugutov potential, which is designed to suppress crystallization [8]. Periodic boundary condition is applied in the simulation to prevent particle evaporation, and the timestep is set to 0.005. Next, we perform MD simulation using Langevin Dynamics [13] with friction coefficient 1. Starting from simple cubic initial configurations, the system equilibrates at a high initial temperature of $T_0$=3 and then cools to a desired sampling temperature $T_s$, and maintain the temperature to calculate the static structure factor. We use a wide range of $T_s$ covering the liquid, supercooled, and glass states. We use reduced units where the Boltzmann constant is 1. The selection of initial temperature is based on Figs. 8 and 9 in reference [12], where we find that the original Dzugutov system with a density of 0.87 is fluid at that temperature. After equilibrating at the initial temperature for 20000 timesteps, the system is cooled down using the cooling schedule $T=T_0e^{-0.005t}$ until $T_s$ is reached. Then, we continue to perform constant temperature simulation for 1000 timesteps to calculate the structure factor, which is defined as

$$\mathrm{S}(\mathbf{q}) = \frac{1}{N}\left|\sum_{j=1}^{N} \mathrm{e}^{-i\mathbf{q}\cdot\mathbf{R}_j}\right|^2. \tag{1}$$

To calculate $\mathrm{S}(\mathbf{q})$ accurately, we perform the above simulations 1000 times for each $T_s$. After excluding any crystallized trajectory detected by $q_8$ (detailed below), there are at least 999 trajectories remaining for each $T_s$. We use the part of these trajectories at $T_s$ to calculate the average structure factor. Specifically, to calculate $\mathrm{S}(\mathbf{q})$, the increment of each component of the wave vector $\mathrm{q_x, q_y, q_z}$ is $\Delta\mathrm{q}=\frac{2\pi}{L}$, where $L$ is the side length of the simulation box. For disordered systems, we simplify $\mathrm{S}(\mathbf{q})$ as a function of the modulus of the wave vector, $\mathrm{q}=|\mathbf{q}|$. We average the structure factors of $\mathbf{q}$ with the same modulus, and finally obtain the average structure factor $\mathrm{S}(\mathrm{q})$. The calculated $\mathrm{S}(\mathrm{q})$ of original Dzugutov system at each $T_s$ serve as the target structure factors for the next section.

Lastly, we implement energy minimization on the last frame of each trajectory using the FIRE algorithm [14,15], which can be considered as quenching the systems' temperature to zero. This enables determining which systems have partially crystallized using $q_8$ order parameter.

Based on the observation of the $q_8$ in our study and the cooling simulations in references [12,16], we find that the original Dzugutov fluids at our density only form Body-Centered Cubic (BCC) crystals. We therefore use $q_8$ of the nearest 14 neighbors to determine the presence of crystallization. Our observations find that a good criterion for crystallization is whether there are more than 9 particles with $q_8$ exceeding 0.4. We use this criterion to exclude crystallized trajectories. If a final configuration has partially crystallized, the complete trajectory is discarded in calculating $\mathrm{S}(\mathrm{q})$.

To determine the onset temperature of glass dynamics, we also calculate the incoherent intermediate scattering function $\mathrm{F_s}(\mathbf{q},\mathrm{t})$ [3,17], which is defined as

$$\mathrm{F_s}(\mathbf{q},\mathrm{t}) = \frac{1}{N}\langle\sum_{\mathrm{j}}^{N} \mathrm{e}^{\mathbf{i}\mathbf{q}\cdot\left(\mathbf{r_j}(\mathrm{t})-\mathbf{r_j}(0)\right)}\rangle, \tag{2}$$

where $N$ is the number of particles, $\mathbf{q}$ is at the first peak of the static structure factor, $\mathrm{t}$ is the time, and $\langle\ldots\rangle$ denotes ensemble average. When the decay of $\mathrm{F_s}(\mathbf{q},\mathrm{t})$ with time no longer exhibits an exponential form but instead begins to develop a plateau, the system exhibits glass dynamics behavior [3,17]. In our study, we determine the glass dynamics onset temperature by observing at which temperature does $\mathrm{F_s}(\mathbf{q},\mathrm{t})$ begin to develop a plateau.

To calculate $\mathrm{F_s}(\mathbf{q},\mathrm{t})$, we perform 100 simulations on each $T_s$ again, since these simulations are much longer than the ones for calculating $\mathrm{S}(\mathbf{q})$. Here we increase the time for constant temperature simulation at temperature $T_s$ to 200000 timesteps, while keeping other conditions unchanged. After ruling out crystallization, there are at least 44 trajectories remaining for each $T_s$. We take a sample every 10 timesteps, and collect 10 pairs of frames per trajectory, to calculate $\mathrm{F_s}(\mathbf{q},\mathrm{t})$ for $\mathrm{t} < 0.5$. We then switch to collecting a sample every 1000 timesteps to calculate $\mathrm{F_s}(\mathbf{q},\mathrm{t})$ for larger $\mathrm{t}$. For each $T_s$, we draw one $\mathrm{F_s}(\mathbf{q},\mathrm{t})$ curve.

*gzhang37@cityu.edu.hk

The above discussion applies to the original Dzugutov system. To calculate S(q) and $F_s(\mathbf{q},t)$ of modified Dzugutov system [9], we keep the protocol unchanged except for three differences. First, we change the interaction between particles to modified Dzugutov potential. Second, we change the friction coefficient in Langevin dynamics to 10 [18]. Third, we skip energy minimization. In all simulations, we have never observed crystallization, so we don't need energy minimization to accurately calculate $q_8$. All trajectories can be used to calculate S(q) and $F_s(\mathbf{q},t)$.

### B. Optimization

We apply an improved version of the algorithm designed in reference [19] to generate configurations that fit target structure factors. The original algorithm starts with a series of random initial configurations and adjusts particle positions in all configurations simultaneously to minimize the mean squared error (MSE) between their average structure factor and the target [19]:

$$\mathrm{MSE} = \frac{1}{N_q} \sum_{|\mathbf{q}|<q_{max}} \left( \frac{1}{N_c} \sum_{i=1}^{N_c} S_i(\mathbf{q}) - S_{target}(\mathbf{q}) \right)^2, \quad (3)$$

where $N_q$ is the number of $\mathbf{q}$ shorter than $q_{max}$, $N_c$ is the number of configurations, and $S_i(\mathbf{q})$ is the structure factor of the ith configuration. The algorithm can accept anisotropic targets, but here we target angular-averaged S(q) from MD simulations. For each $q_{max}$, we choose a suitable value of $N_c$, such that the total degrees of freedom ($N_c \times N \times 3$) is at least 10 times larger than the number of constrains ($N_q$). This ensures that the optimization problem is underconstrained, which improves efficiency [19].

In our study, we replace the LBFGS [20] optimization algorithm used in reference [19] with Adam [21], a variant of the stochastic gradient descent (SGD) algorithm in machine learning. In each step the optimizer only uses a small sample of $\mathbf{q}$ to estimate the gradient direction. The sample size, called "batch size" in SGD terminology, is 256 in our study. The step sizes (learning rate) starts form $3 \times 10^{-4}$ and is reduced by half whenever MSE does not improve for 5 epochs. We carry out 10000 epochs of optimizations. As detailed in the supplementary material, less epochs would not reveal the onset temperature of glass dynamics. We compare the optimized structure factors and targets in the supplementary material.

It is important to note that only the target structure factors are used as input to the algorithm, so the algorithm does not know the configurations sampled from MD simulations. The input structure factor has a cutoff wavevector, $q_{max}$. The information of structure factor with q exceeding $q_{max}$ will not be provided to the optimizer. Choosing a proper $q_{max}$ is important in our study, since increasing $q_{max}$ will significantly increase the computation cost, while reducing $q_{max}$ reduces the information accessible to the algorithm. We need to find a suitable range for $q_{max}$, so we select a series of $q_{max}$ to perform optimizations. After that, we plot $\lg(\mathrm{MSE}) = \log_{10}(\mathrm{MSE})$ versus the temperature. To reduce fluctuations of the curves, we perform multiple optimization runs for each structure factor and take the median MSE. Finally, we observe the characteristics of lg(MSE)-Temperature curve and compare it with $F_s(\mathbf{q},t)$.

## III. RESULTS

The calculated structure factors are presented in the supplementary material. Here we examine the lg(MSE)-Temperature curves and $F_s(\mathbf{q},t)$.

Firstly, we study the original Dzugutov system. We set $q_{max}$ to 15, 17, 19, 21, and 23. For each $q_{max}$, we perform optimizations and plot the resulting lg(MSE)-Temperature curves. The curves for $q_{max}$=15 and $q_{max}$=17 only show negative correlations between lg(MSE) and temperature without any clear kink, so they are presented in the supplementary material. The results for $q_{max}$=19, 21, 23, shown in Fig. 1, are flat in high temperature regions but are steep in low temperature regions. There is a clear non-differentiable point at T=0.7 or 0.8 in each curve. This shows that before the onset of glass dynamics, the structure factors are easy to realize. As the temperature drops, MSE increases, which signifies the difficulty in realizing such S(q). In addition, we observe in Fig. 2 that $F_s(\mathbf{q},t)$ of this system starts to deviate from exponential decay and develops a plateau when the temperature is around 0.8. The consistency between the kinks in lg(MSE) curves and the plateau appearing temperature of $F_s(\mathbf{q},t)$ demonstrates that the onset of glass dynamics can be detected by the median MSE achieved in our optimizations.

*gzhang37@cityu.edu.hk

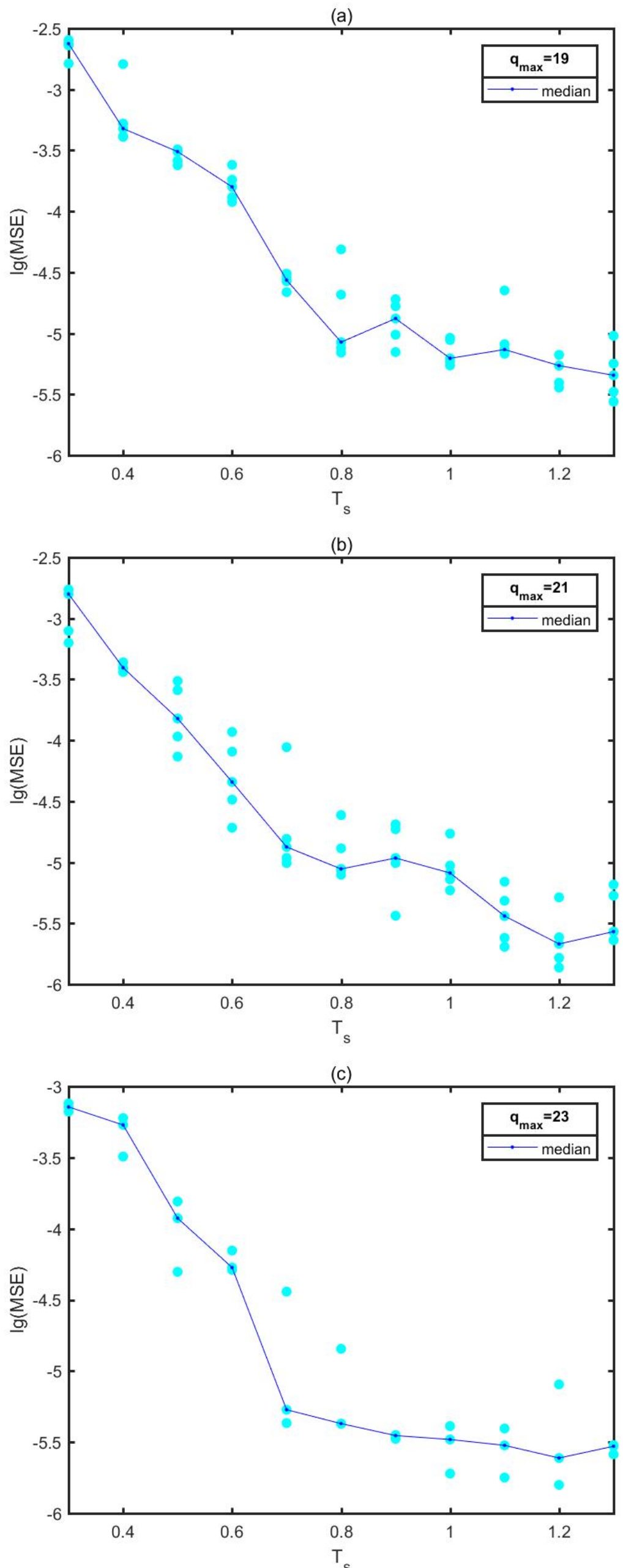

FIG 1. The lg(MSE) -Temperature curves of the original Dzugutov system. Light blue dots represent individual runs. The solid lines connect median values at each temperature.

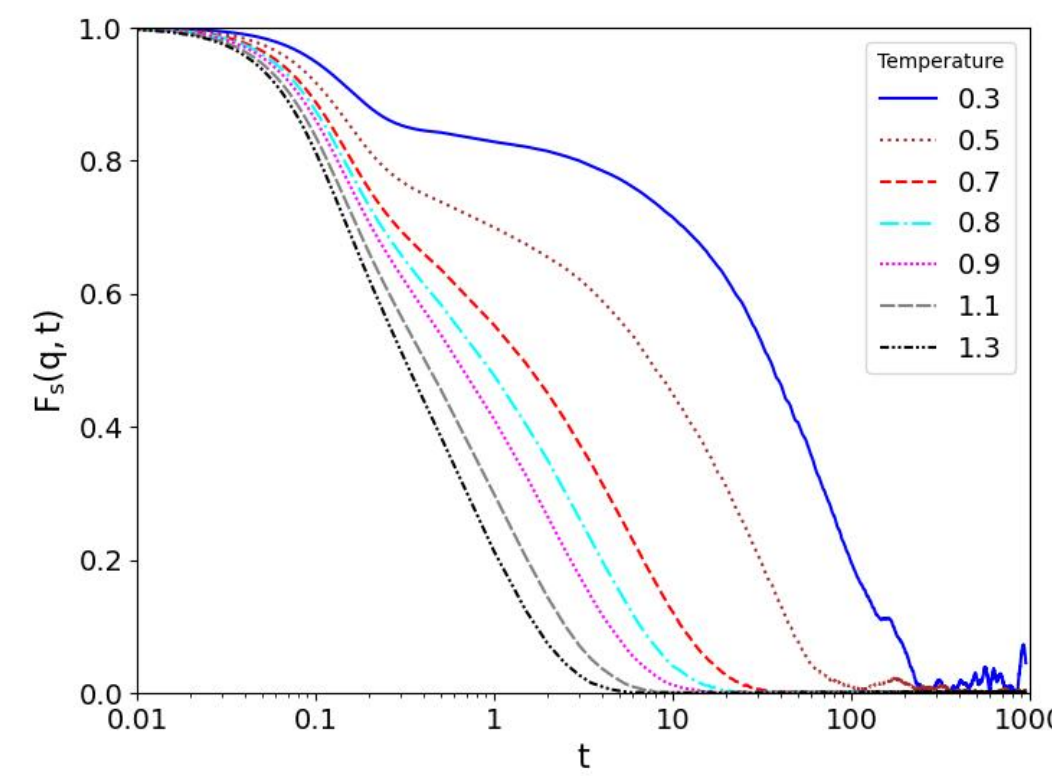

FIG 2. The incoherent intermediate scattering function $F_s(\mathbf{q}, t)$ of the original Dzugutov system.

The curve of $q_{max}$=21 are flat immediately to the right of the kink, but continues to drop for $T_s > 1$. However, the results for $q_{max}$=20.5 and $q_{max}$=21.5, shown in Fig. 3, exhibits no such drop. For $T_s < 1$ these two curves are similar to the curve of $q_{max}$=21. This shows that whether the curve decreases at the highest temperature is sensitive to the selection of $q_{max}$, and therefore cannot be relevant to glass dynamics.

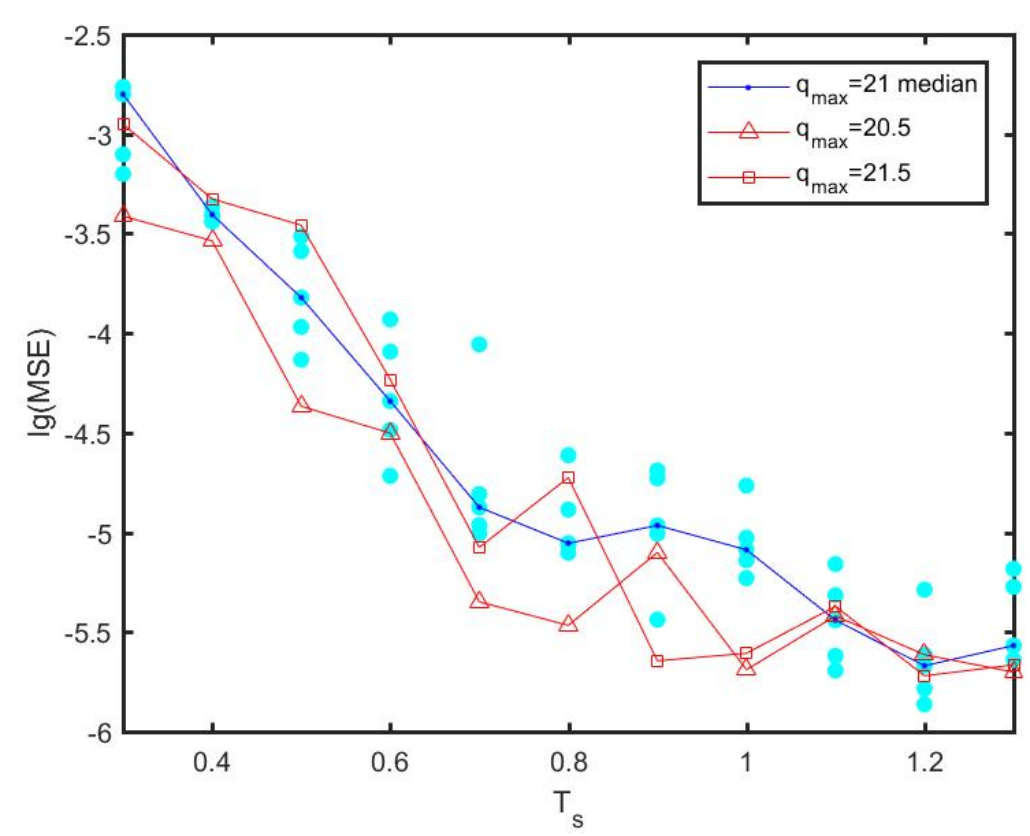

FIG 3. The lg(MSE) -Temperature curves of the original Dzugutov system when $q_{max} = 21, 20.5, 21.5$. Light blue dots represent individual runs for $q_{max} = 21$. For $q_{max} =$ 20.5 and 21.5, each target structure factor is optimized only once.

In order to reduce the influence of the above phenomenon on determining the glass dynamics onset temperature and reduce computational costs, a reasonable algorithm is to average the lg(MSE) curves at slightly different $q_{max}$ values. Starting from $q_{max}$=15, one can gradually increase $q_{max}$ and draw one lg(MSE) -Temperature curve for each $q_{max}$. If three consecutive curves exhibit similar behavior and clear kinks, one take the average of these three curves, and the averaged mean(lg(MSE))-Temperature curve would clearly determine the glass dynamics onset temperature. For our original Dzugutov system, such an algorithm would select $q_{max}$ =19, 21, 23 for calculating the mean(lg(MSE))-Temperature curve, shown in Fig. 4. This curve clearly shows that the onset temperature is 0.7.

*gzhang37@cityu.edu.hk

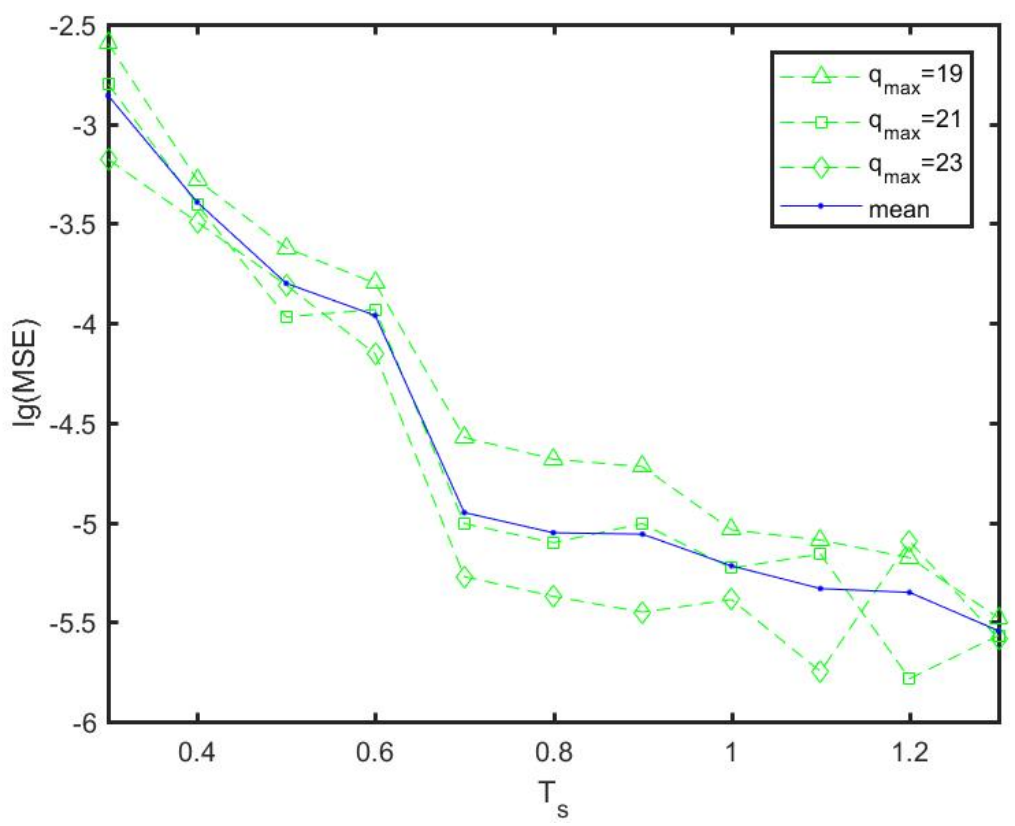

FIG 4. The dashed lines are $\lg(\mathrm{MSE})$-Temperature curves of the original Dzugutov system with three different $q_{max}$. The solid line is their average.

We then study the modified Dzugutov system and find similar results. We experimented with $q_{max}$=15, 17, 19, 21, 23, 25, and 27. When $q_{max}$ is less than 23, there are no clear kinks in the curves. Part of these results are presented in the supplementary material. As shown in Fig. 5, the $\lg(\mathrm{MSE})$-Temperature curves of $q_{max}$=23 and 25 both have kinks at around T=0.65. The kink of $q_{max}$=27 is not obvious, and we suspect that the reason may be due to insufficient epochs of optimization. After increasing the epoch from 10000 to 15000, the curve of $q_{max}$=27 also exhibits a kink at around 0.65. Meanwhile, the $F_s(\mathbf{q}, t)$ develops a plateau when the temperature is around 0.6, as shown in Fig. 6. As with the original Dzugutov system, our algorithm clearly detects the onset of glass dynamics.

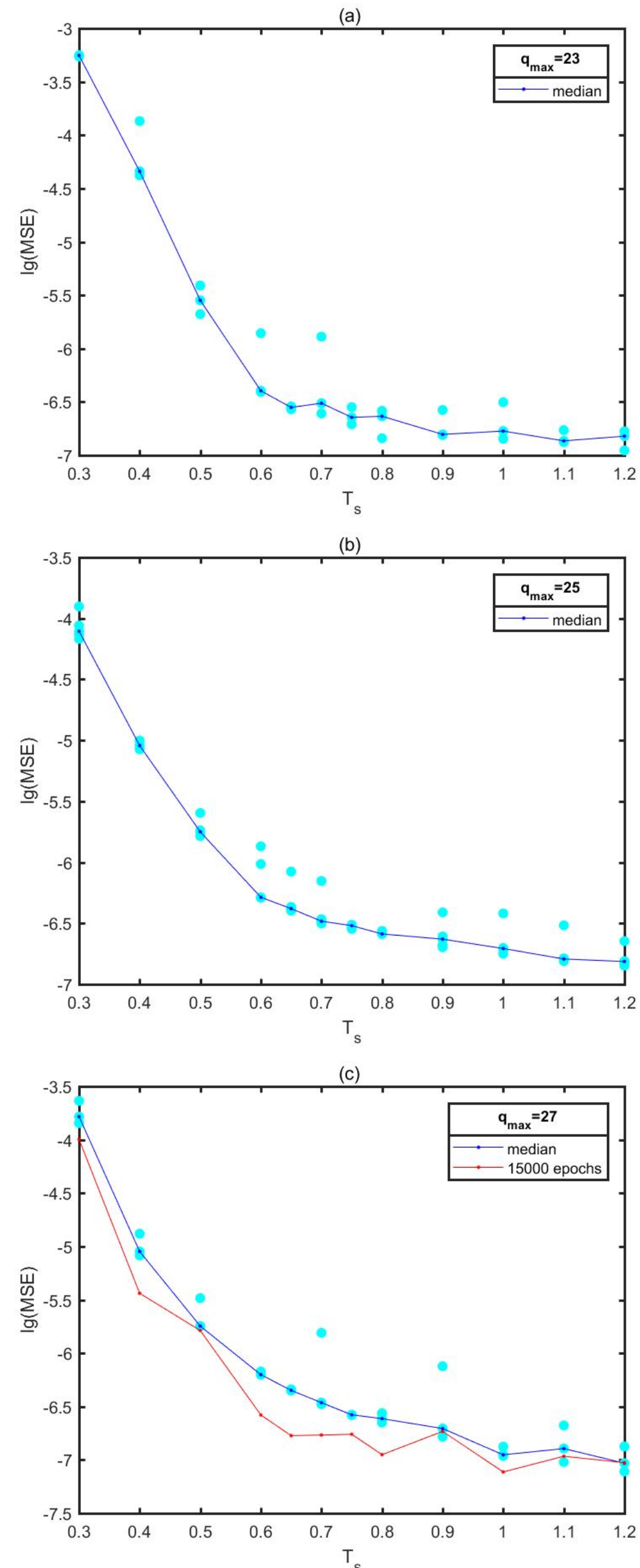

FIG 5. The $\lg(\mathrm{MSE})$-Temperature curves of the modified Dzugutov system. Light blue dots represent individual runs. The solid lines connect median values at each temperature. The read line in figure (c) is achieved after optimizing for 15000 epochs instead of 10000.

*gzhang37@cityu.edu.hk

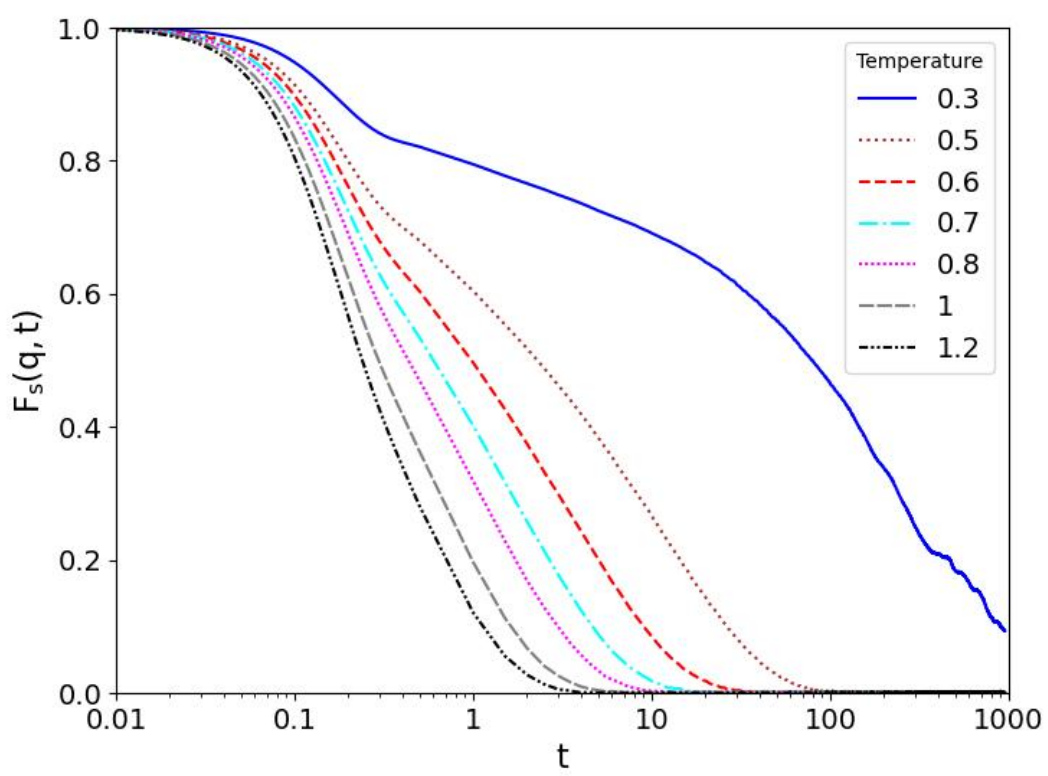

FIG 6. The incoherent intermediate scattering function $F_s(\mathbf{q},t)$ of the modified Dzugutov system.

Similar to the original Dzugutov system, we can average over three different $q_{max}$ values to obtain a smoother mean(lg(MSE))-Temperature curve with a kink at 0.65, shown in Fig. 7.

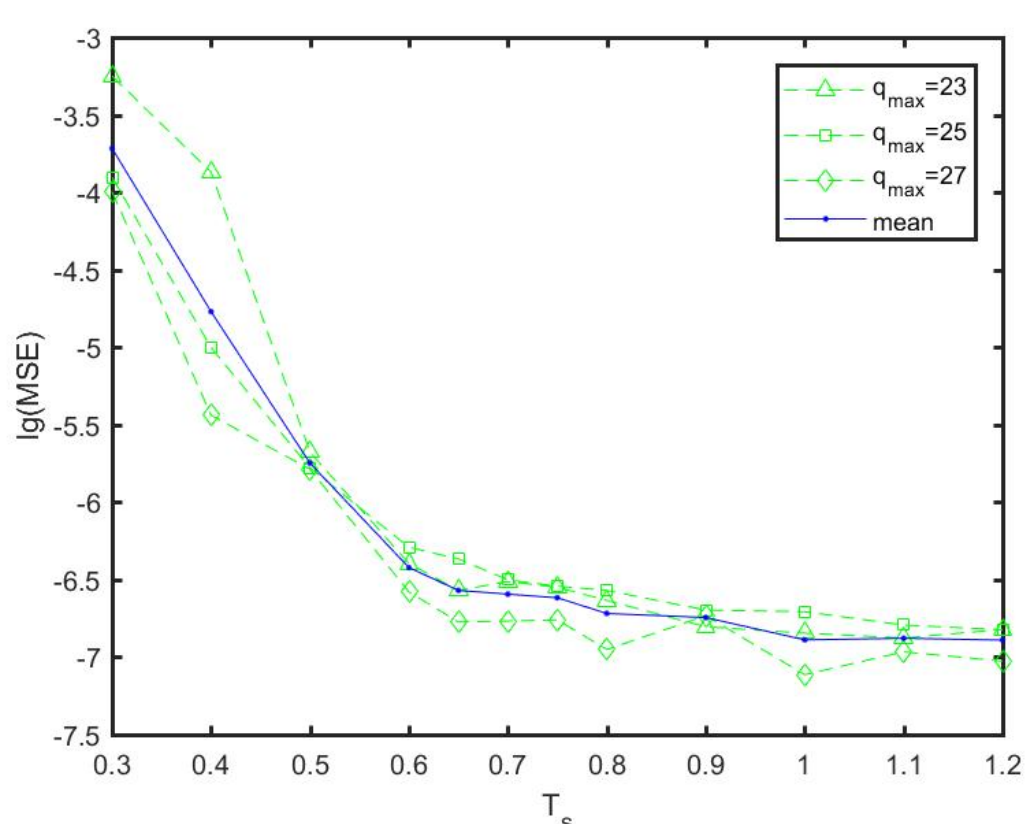

FIG 7. The dashed lines are $\lg(\mathrm{MSE})$-Temperature curves of the modified Dzugutov system with three different $q_{max}$. The solid line is their average.

## IV. CONCLUSIONS AND DISCUSSION

We applied stochastic gradient descent in realizing $S(q)$ of original and modified Dzugutov systems at various temperatures. We observed that the difficulty in optimization, measured in lg(MSE), exhibits a kink at the glass dynamics onset temperature. This indicates that as the temperature changes, the changes in structure factors are quantitatively small but are qualitatively important. Based on this, we have developed an algorithm to determine the glass dynamics onset temperature of single-component disordered systems.

While we did not explicitly map the connectivity of the MSE landscape (e.g., via Nudged Elastic Band methods), the divergence in optimization outcomes is consistent with topological transitions similar to the clustering transition in constraint satisfaction problems [22]. At high temperatures, the reconstruction converges to the global minimum. This indicates that the MSE landscape is dominated by a broad, convex basin of attraction. Conversely, at low temperatures, the algorithm arrests at non-zero MSE values, even though MSE=0 solutions exist [ $S(q)$ at low temperature is still realizable]. This suggests a rugged MSE landscape, as depicted in Fig. 8. Such a transition, if exists, could explain why glass dynamics are slow: the part of the configuration space corresponding to the low temperature $S(q)$ (with MSE=0) is fragmented, preventing the free evolution of the system.

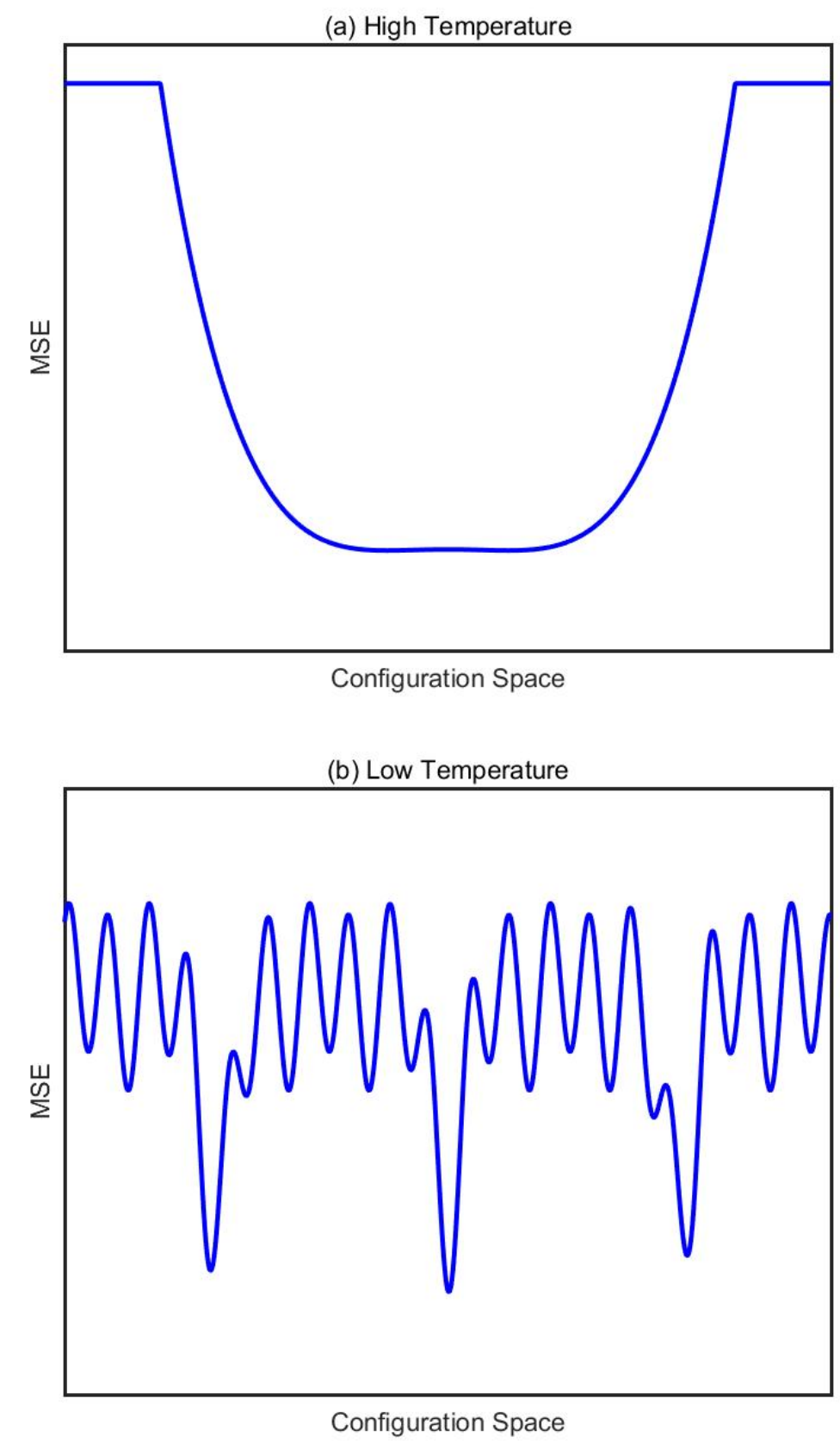

FIG 8. Schematics of the MSE landscape for the target structure factors at high and low temperatures, respectively. The landscape is smooth with a single global minimum at high temperatures, but develops many local minima at lower temperatures.

A well-known open problem in statistical mechanics is the pair-statistics realizability problem [23, 24]: Given a pair correlation function or a structure factor, is it realizable at a specific number density? Our results suggest that it is informative to further subdivide realizable pair statistics into two categories: those with a smooth MSE landscape (easily realizable), and those with a rugged landscape

*gzhang37@cityu.edu.hk

(difficult to realize). As our results demonstrated, difference in the difficulty of realization can lead to differences in physical properties.

It would be worthwhile to continue improving our algorithm in the future. First, when optimizing with a target $S(q)$, the cutoff wavevector $q_{max}$ usually cannot be too small, and the minimum value of $q_{max}$ for observing the kinks varies for different systems. We find that such a minimum $q_{max}$ is 19 for the original Dzugutov system and 23 for the modified Dzugutov system. Since the modified Dzugutov potential has a softer repulsive core than the original, we suspect that our method is more suitable for stiffer potentials. However, the specific reason still needs to be discovered in further studies.

Second, we have not identified the factors that affect the fluctuations of the $\lg(\mathrm{MSE})$-Temperature curves. Our results show that some curves have small fluctuations while others have large fluctuations. Currently we can optimize multiple times and take the median MSE to combat fluctuations, but the computational cost also multiplicates. We suspect that the parameters in the optimization algorithm may benefit from further tuning, as the current parameter settings are mainly aimed at accelerating convergence.

Lastly, we plan to extend our algorithm to other complex glass-forming systems. We hope to collaborate with experimental scientists to verify whether our method is applicable for actual glass samples, as measuring $S(\mathbf{q})$ in experiments is easier than measuring $F_s(\mathbf{q}, t)$. To facilitate this, we plan to extend our method to multi-component systems and more complex interactions (e.g., between organic molecules).

## SUPPLEMENTARY MATERIAL

The supplementary material contains structure factors calculated from MD simulations, comparison between $S(q)$ targets and optimized configurations, the $\lg(\mathrm{MSE})$-Temperature curves when $q_{max}$ is too small and when the optimization is not finished.

## ACKNOWLEDGMENTS

We thank Massimo Pica Ciamarra for helpful discussions. The work described in this paper was fully supported by a grant from City University of Hong Kong (Project No. 9610532).

*gzhang37@cityu.edu.hk

*gzhang37@cityu.edu.hk

# Supplementary Material:
# Structure factor realizability reveals the glass-dynamics onset temperature

Bowen Duan[1] and Ge Zhang[1]

[1]Department of Physics, City University of Hong Kong, Hong Kong, China

## 1. Structure factors calculated from MD Simulations

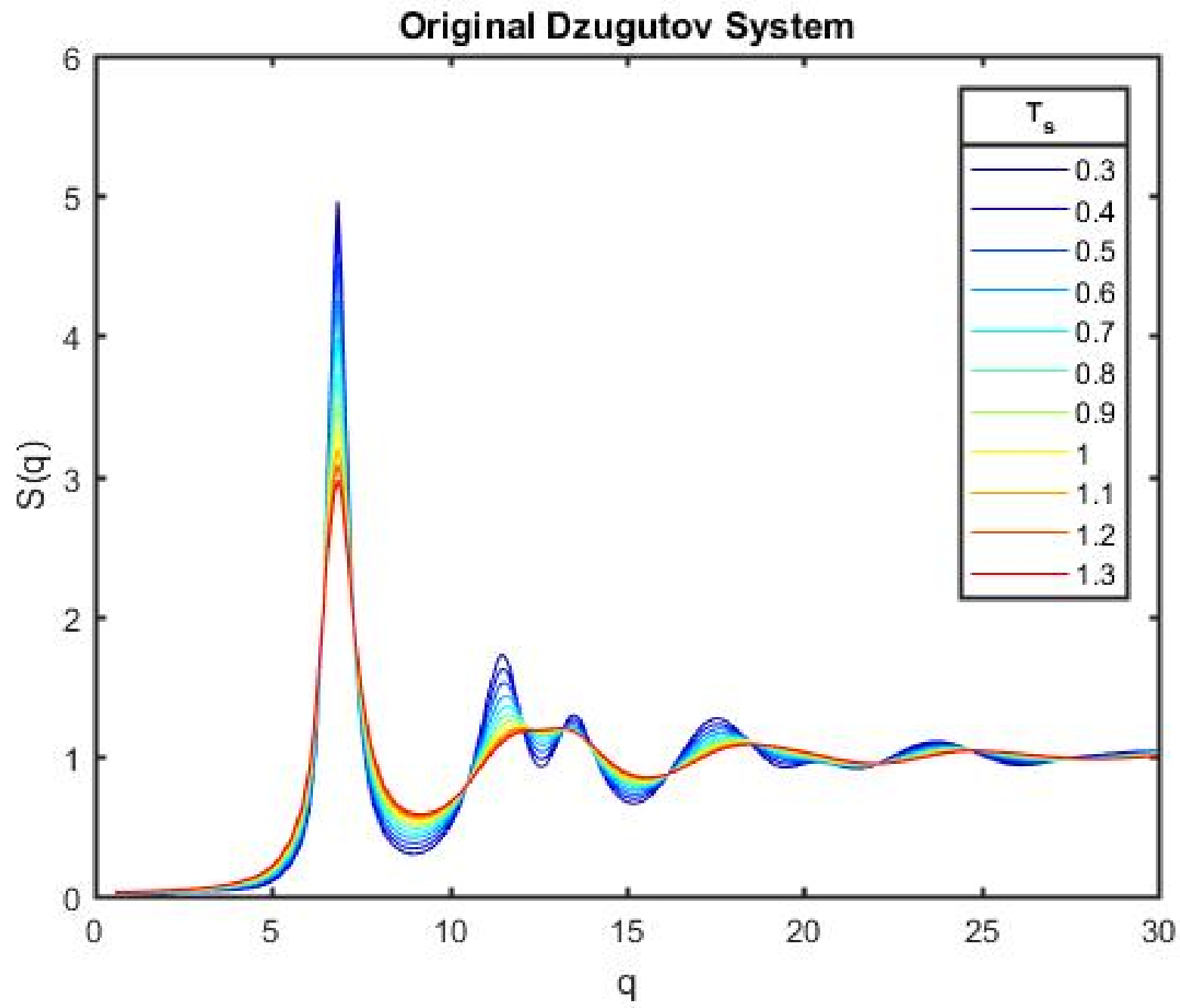

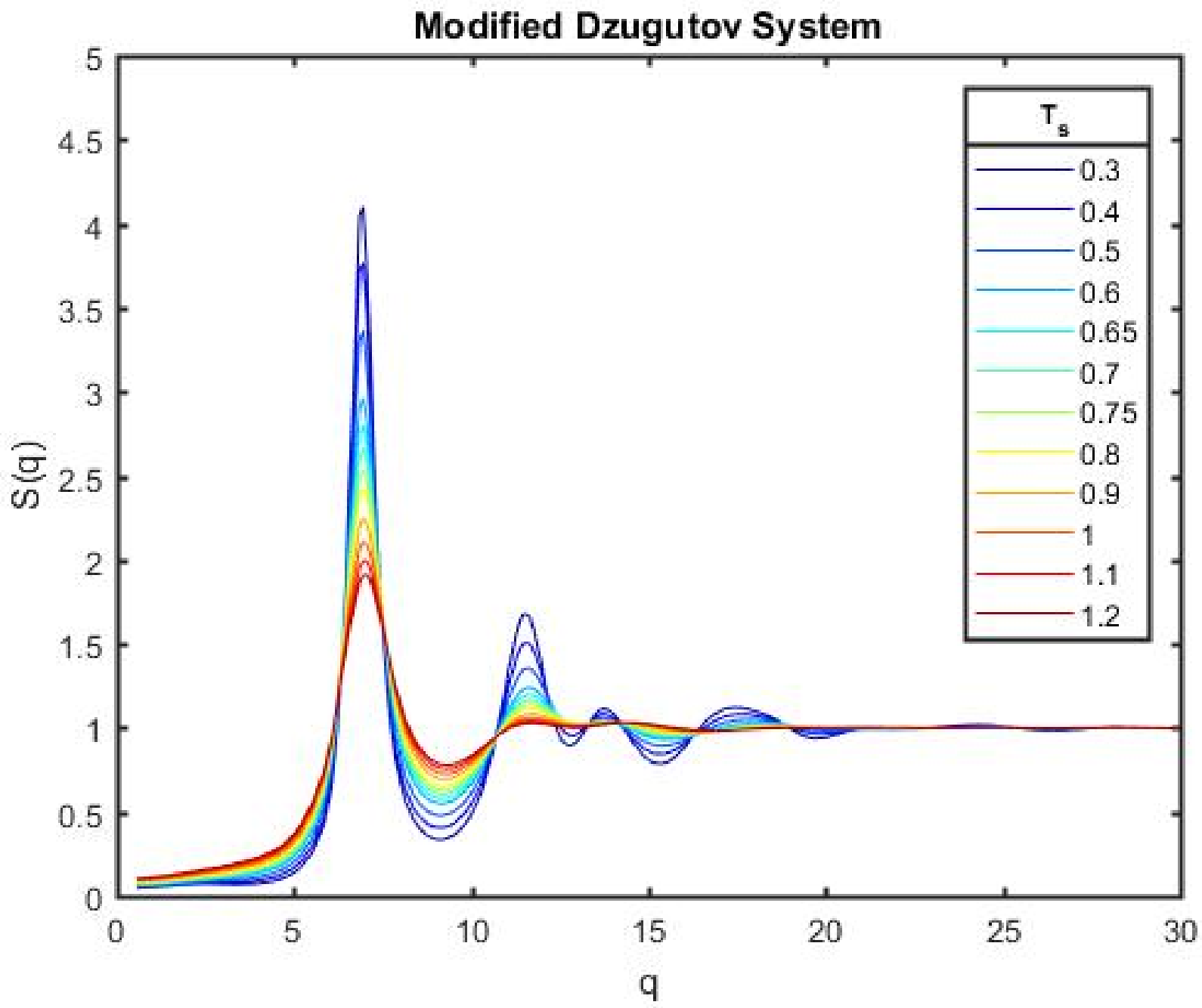

## 2. Comparison of the optimized structure factors and targets

Here we present a comparison between the optimized structure factors and the targets at low and high temperatures. For original Dzugutov system, the $q_{max}$ of the input structure factor is 19 and the number of initial random configurations is 300. The optimized structure factor is the average structure factor of 300 final output configurations. For modified Dzugutov system, the $q_{max}$ is 23 and the number of configurations is 450.

Original Dzugutov System

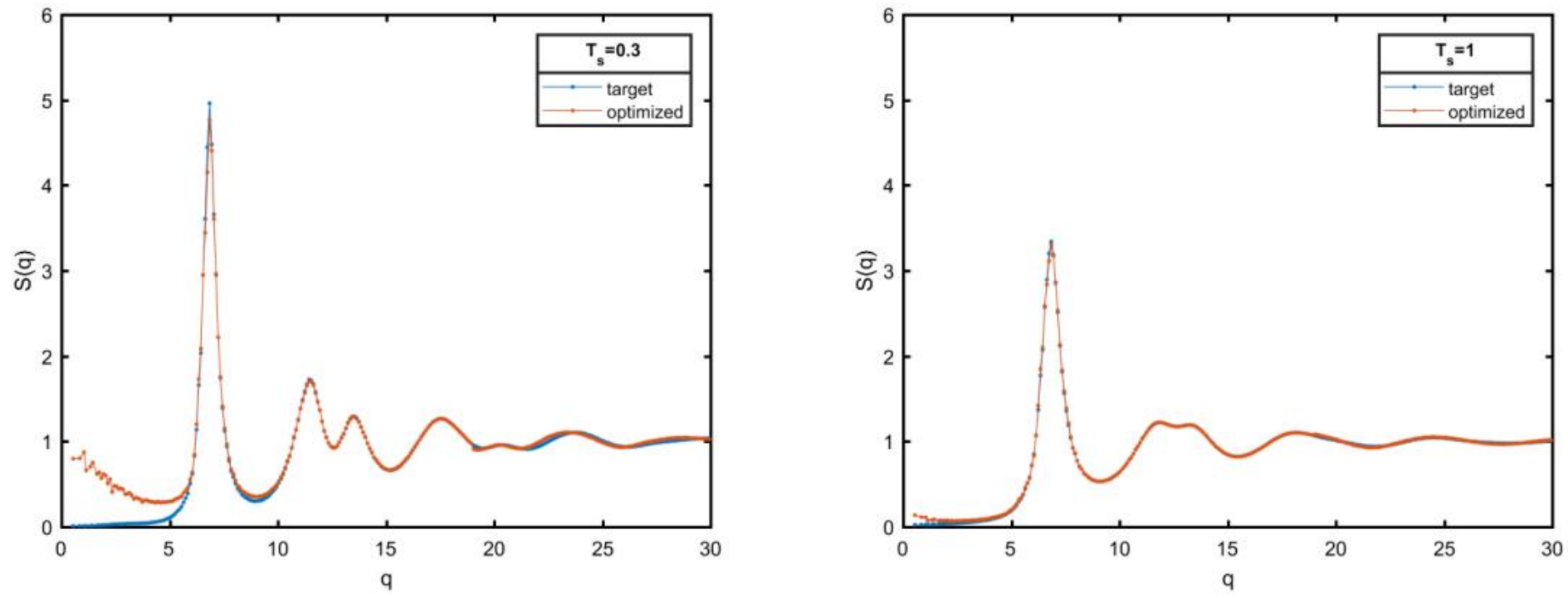

Modified Dzugutov System

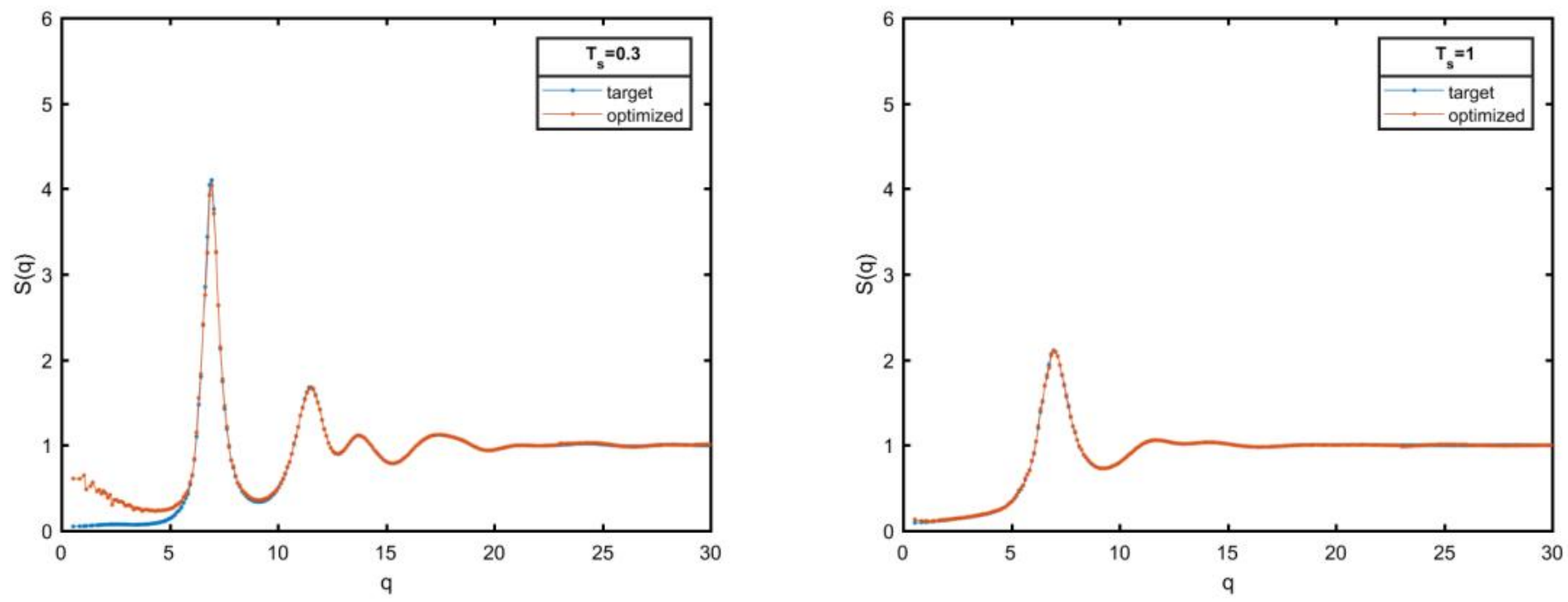

## 3. The lg(MSE)-Temperature curves with smaller $q_{max}$

As detailed in the main test, the lg(MSE)-Temperature curve exhibits non-differentiable point only if $q_{max}$ is sufficiently high. Here we present such curves with lower $q_{max}$.

Original Dzugutov System

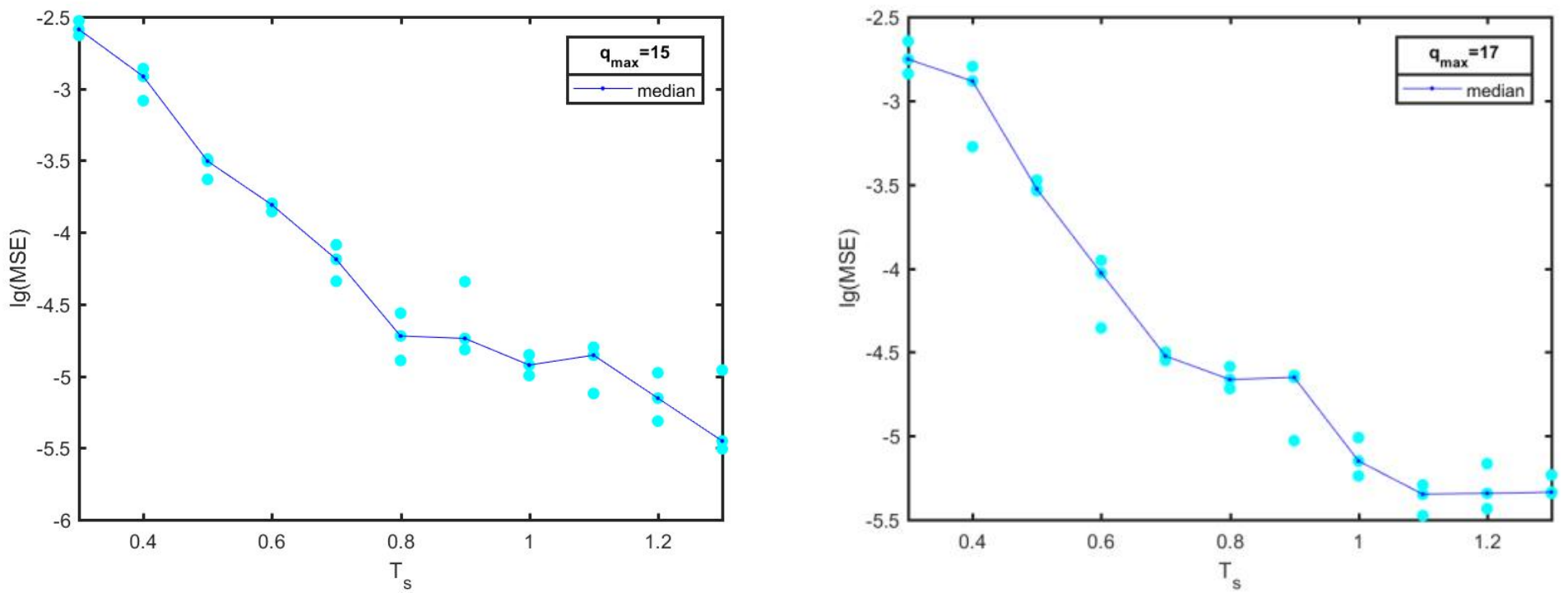

Modified Dzugutov System

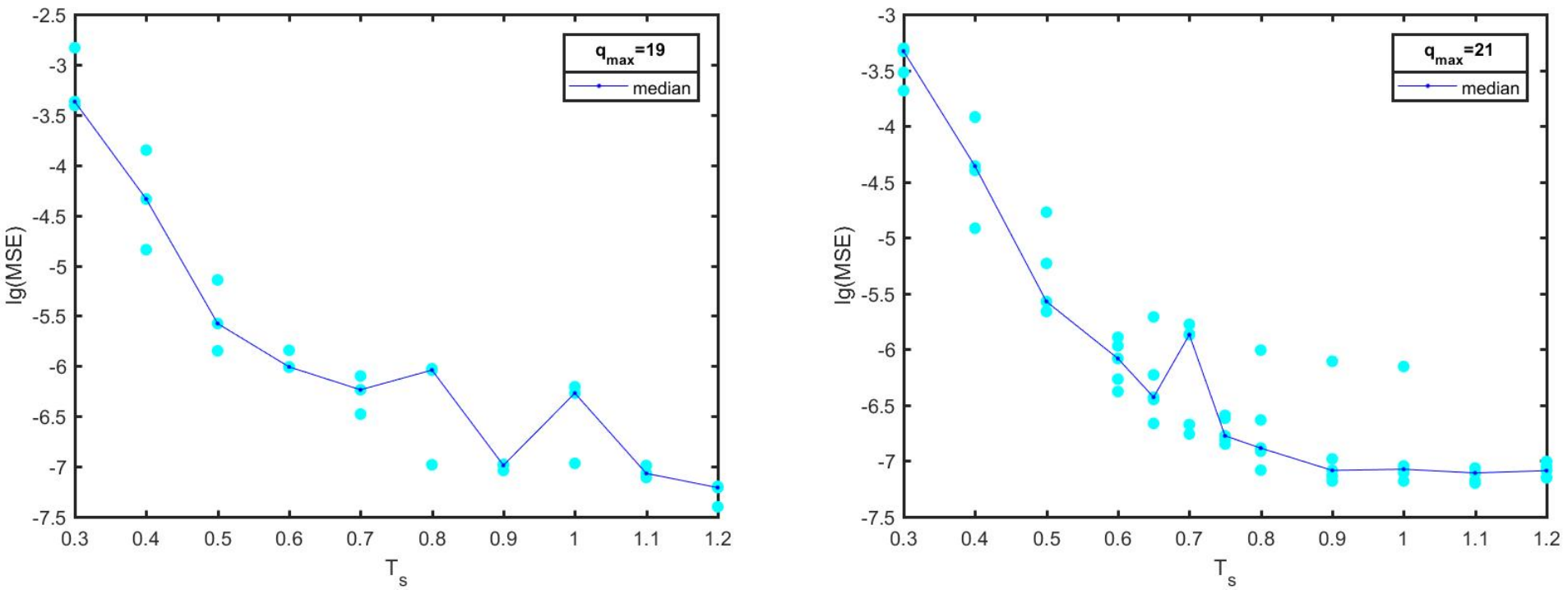

## 4. The behaviors of the lg(MSE)-Temperature curves during the optimization process

Here we present lg(MSE)-Temperature curves at different stages of a single optimization of original Dzugutov system with $q_{max} = 23$. The curves are initially smooth and then develop non-differentiable point.

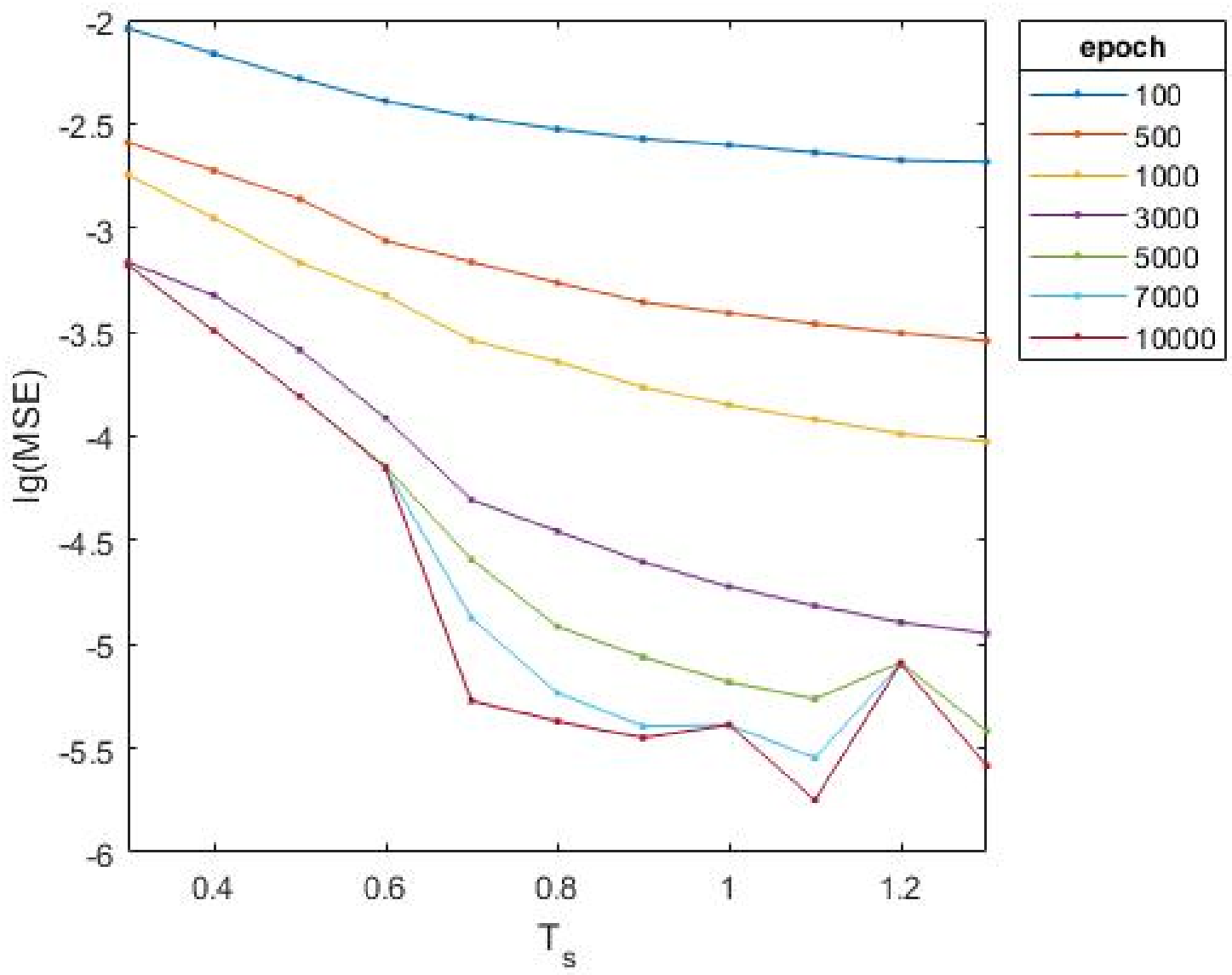